\documentclass[12pt,letterpaper]{article} 

\usepackage[dvips]{graphicx}
\usepackage{fullpage}
\usepackage{sectsty}
\sectionfont{\large}

\begin{document}

\begin{center}
\Large\bfseries
Model uncertainties on limits for quantum black hole production
in dijet events from ATLAS
\end{center}
\bigskip

\begin{center}
\small
Douglas M. Gingrich$^{1,2}$ and Krishan Saraswat$^{1}$\\

\bigskip

\textit{$^1$Centre for Particle Physics, Department of Physics,
University of Alberta,\\
Edmonton, AB T6G 2E1 Canada}\\  
\textit{$^2$TRIUMF, Vancouver, BC V6T 2A3 Canada}\\ 
{\footnotesize gingrich@ualberta.ca}
\end{center}

\begin{center}
\small \today
\end{center}

\begin{abstract}
We study the model uncertainties on limits for quantum black hole
production in dijet events from ATLAS.
For models that assume a hard-disk cross section, the model uncertainty
on the threshold mass limits is about 5\%.
If the trapped surface calculation is used for the cross section, the
ATLAS mass threshold limits are below 2~TeV for all number of dimensions.
Using the ATLAS data in the context of the Randall-Sundrum type-1 model
gives a threshold mass lower limit of 2.84~TeV.
\end{abstract}

\section{Introduction} \label{sec1}

\begin{figure}[htb]
\centering
\includegraphics[width=8cm]{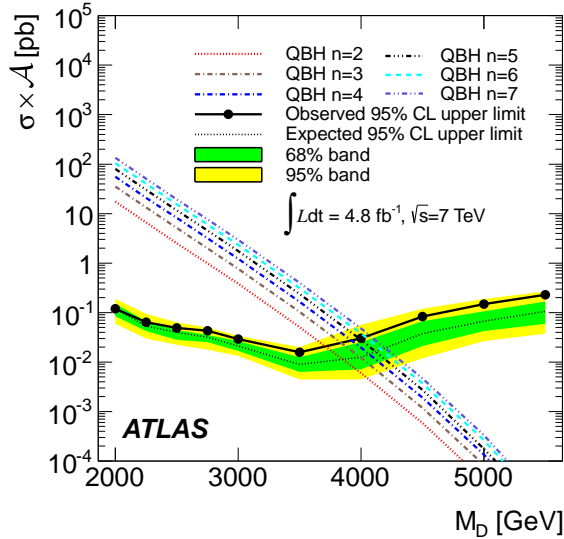}
\caption{The 95\% C.L. upper limits on $\sigma \times \mathcal{A}$ as 
function of the reduced Planck mass $M_D$ of the quantum black hole
models using $F_\chi(m_{jj})$ (black filled circles).
The black dotted curve shows the 95\% C.L. upper limit expected from
Monte Carlo and the light and dark yellow shaded bands represent the
68\% and 95\% contours of the expected limit, respectively.
Theoretical predictions of $\sigma \times \mathcal{A}$ are shown for
various numbers of extra dimensions.
From Ref.~\cite{1210.1718}.}
\label{fig1}
\end{figure}

The ATLAS experiment has set limits on quantum black hole production and
decay to dijets~\cite{Aad:2011aj}.
The analysis has recently been updated to 4.8~fb$^{-1}$ of 7~TeV data~\cite{1210.1718}.
The main result of the new analysis is reproduced in Fig.~\ref{fig1}. 
ATLAS gives the results as a function of the fundamental
higher-dimensional Planck scale\footnote{ATLAS refers to this Planck
scale as the ``reduced'' Planck scale, which is usual reserved for the
four-dimensional Planck scale $\bar{M}_\mathrm{Pl} =
M_\mathrm{Pl}/\sqrt{8\pi}$.} $M_D$. 
The model used by ATLAS equates the value of the Planck scale with the
turn-on, or energy threshold, for quantum black hole production. 
A more model independent interpretation would be to call the $M_D$-axis
in  Fig.~\ref{fig1} the mass threshold for quantum black hole production.
We use this terminology throughout this note.

Several similar models have been proposed to describe the behaviour of
quantum black holes~\cite{Meade:2007sz,Calmet:2008dg,Gingrich:2009hj}.
We study the sensitivity of the ATLAS results to the different models.
The ATLAS results are based on the BlackMax Monte Carlo (MC) event
generator~\cite{Dai:2009by}.
To enable the calculation of difference models we use the {\sc Qbh} MC
event generator~\cite{Gingrich:2009da}\footnote{The use of {\sc Qbh} here
refers to the MC event generator.  
In the ATLAS studies QBH it refers to ``Quantum Black Hole''.}. 

\section{Two-body Branching Ratio} \label{sec2}

Different models for the decay of a quantum black hole to two partons
could give different results for the mass threshold.
Prior to version 2.00.2 of {\sc BlackMax} there was an error in the
two-body branching ratio calculation.
The erroneous formula is still in the manual~\cite{Dai:2007ki} as
Eq.~(23) of that document.  
In this equation, the prefactor should not be raised to an exponent.
This formula also appears in error in the original Meede and Randell
paper~\cite{Meade:2007sz} from which {\sc BlackMax} is based on.
Published results from ATLAS use a version of {\sc BlackMax} with the
corrected formula.

Models for the number of particles produced from the decay of a quantum
black hole use a Poisson distribution 

\begin{equation}
p(n;\nu) = \frac{\nu^n e^{-\nu}}{n!}\, ,
\end{equation}

\noindent
where $n$ is the number observed and $\nu$ is the mean of that
distribution, to describe the probability of different number of final
state particles.   
Different models use this formula in different ways to predict the
probability of a two-body decay.

In {\sc BlackMax}, the two-body branching ratio is given by 

\begin{equation}\label{eq2}
BR = p(0;\nu) + p(1;\nu) + p(2;\nu)\, .
\end{equation}

\noindent
The interpretion is that $p(0;\nu)$ and $p(1;\nu)$ represent processes
in which the black hole does not form, and thus the two incident partons
become the outgoing partons.
Both these terms thus represent a gravitational scattering process using
the classical hard-disk cross section.
The $p(2;\nu)$ term represents a proper two-body decay of a shortlived
quantum black hole.
It is implied by using Eq.~(\ref{eq2}) that all two-body final
states consist of two partons, with no leptons, or gauge bosons
allowed.
Using Eq.~(\ref{eq2}) for the two-body branching ratio reduces the 
limits on the threshold mass by at most 30~GeV from the case of not
including a branching ratio.

In {\sc Qbh}, the two-body branching ratio is given by 

\begin{equation}\label{eq3}
BR = p(1;\nu) / (1 - p(0;\nu))\, .
\end{equation}

\noindent
The interpretation is that we are interested in true decays, not
scattering processes with the classical hard-disk cross section.
Thus $p(0;\nu)$ and $p(1;\nu)$ do not represent physical production or
decays states, and hence are removed from the calculation and the
Poisson distribution is renormalised.
$p(1;\nu)$ represents a proper two-body decay in which one particle is
emitted from the black hole and the remaining black holes state becomes
the second particle.
This is also the interpretation used by the MC event generator
Charybdis2~\cite{Frost:2009cf}. 
Using Eq.~(\ref{eq3}) for the two-body branching ratio reduces the
limits on the threshold mass by at most 120~GeV from the case of not
including a branching ratio, or at most 90~GeV relative to {\sc BlackMax}. 

Not all two-body final states should be consider to give rise to two
jets. 
Based on Hawking emissivities in higher dimensions and enumerating the
number of degrees of freedom of the Standard Model, the probability of
the two final states being both partons is about
64\%~\cite{Gingrich:2009hj}.  
This lowers the limits on the threshold mass by a further 90~GeV.
Thus the {\sc Qbh} model give results at least 180~GeV lower than the ATLAS
results. 

\begin{figure}
\centering
\includegraphics[width=8.1cm]{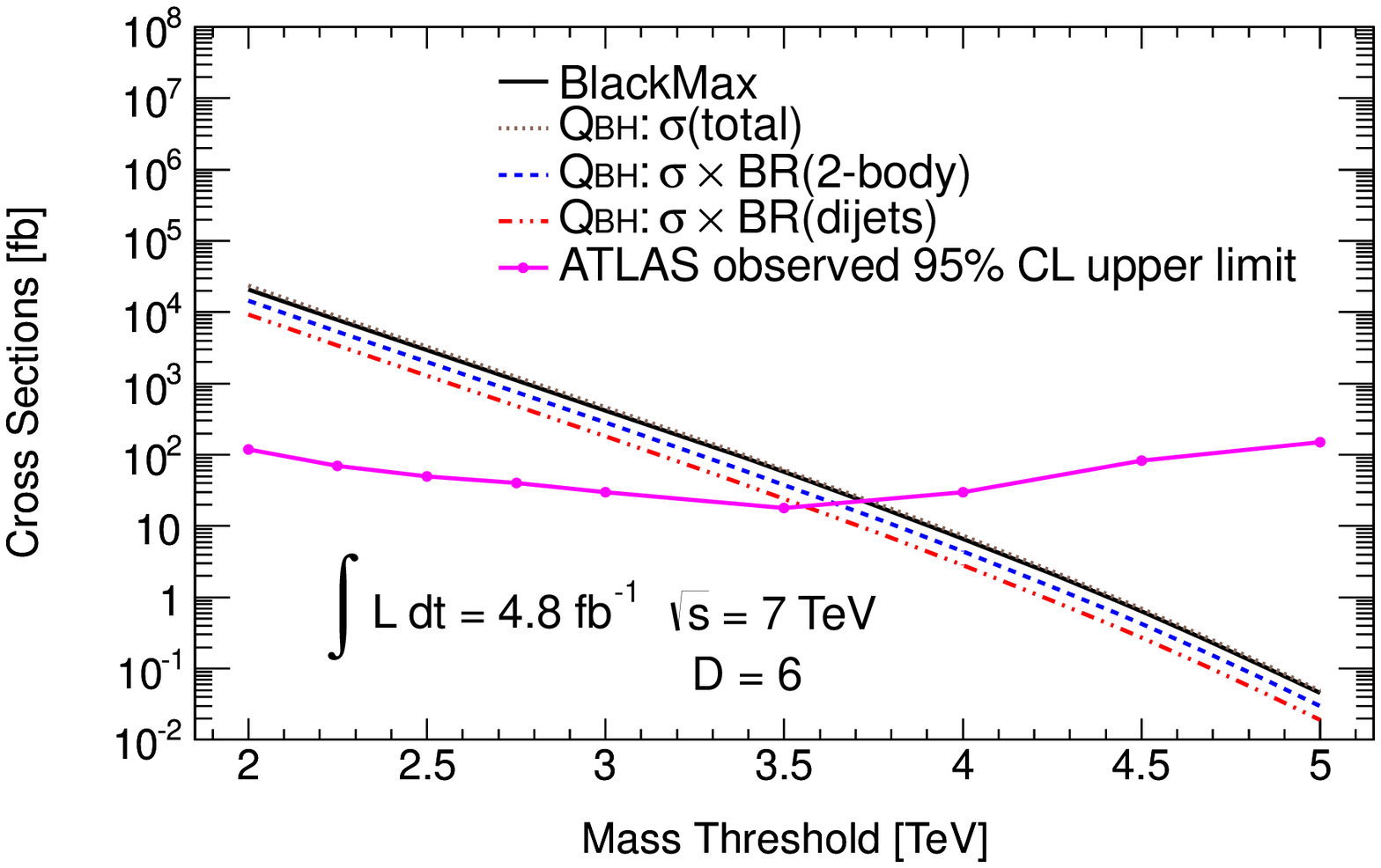}
\includegraphics[width=8.1cm]{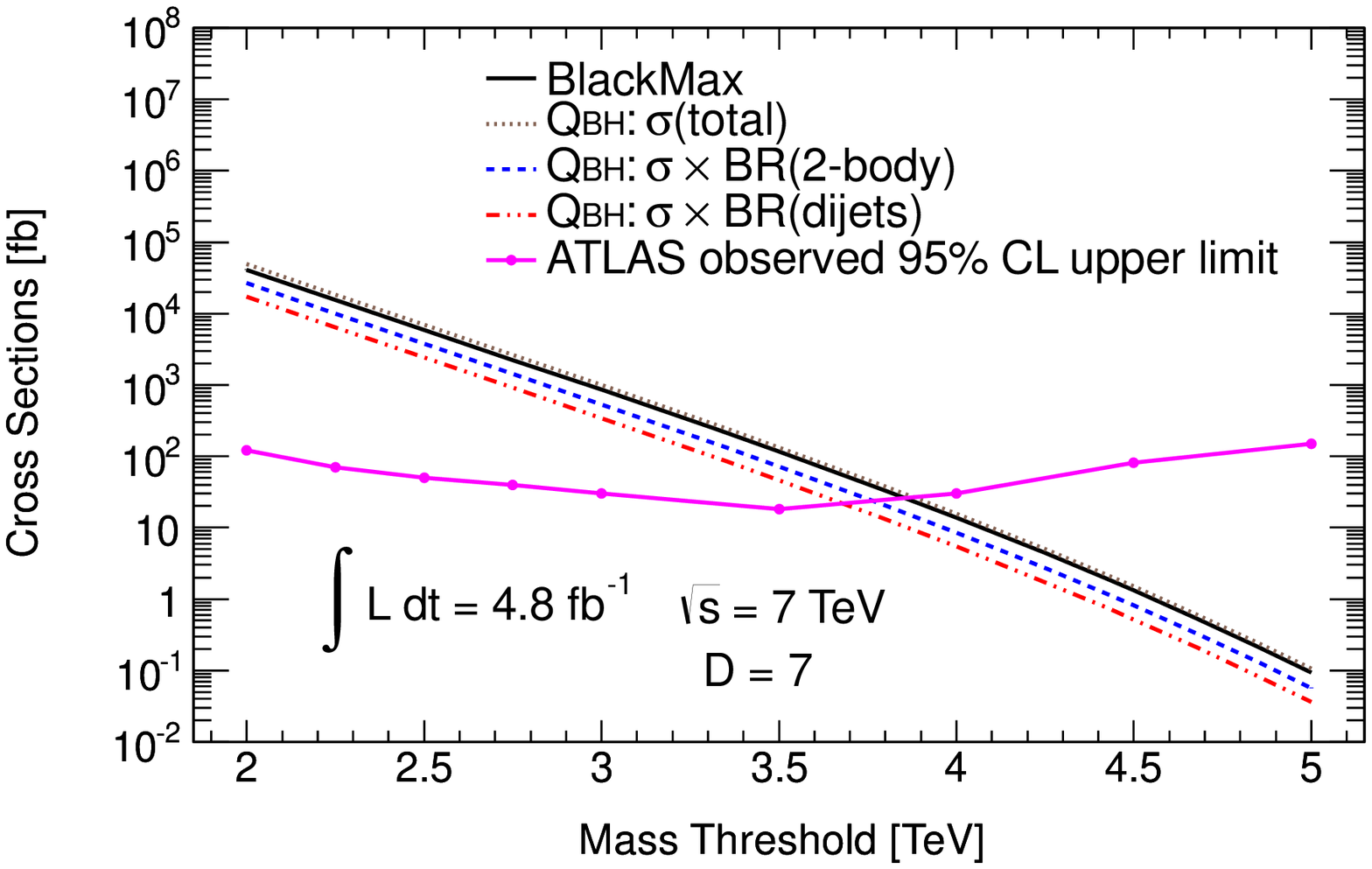}
\includegraphics[width=8.1cm]{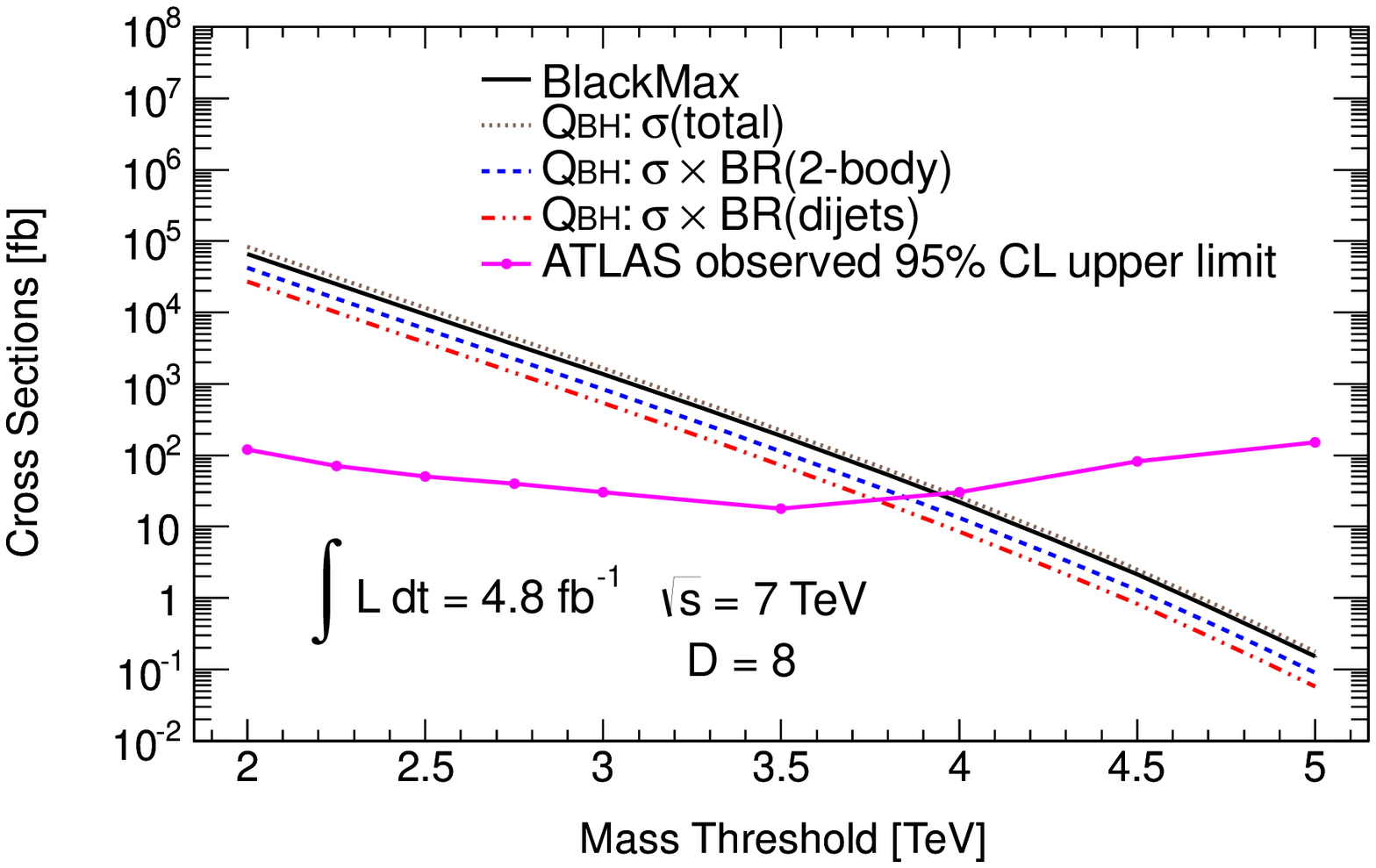}
\includegraphics[width=8.1cm]{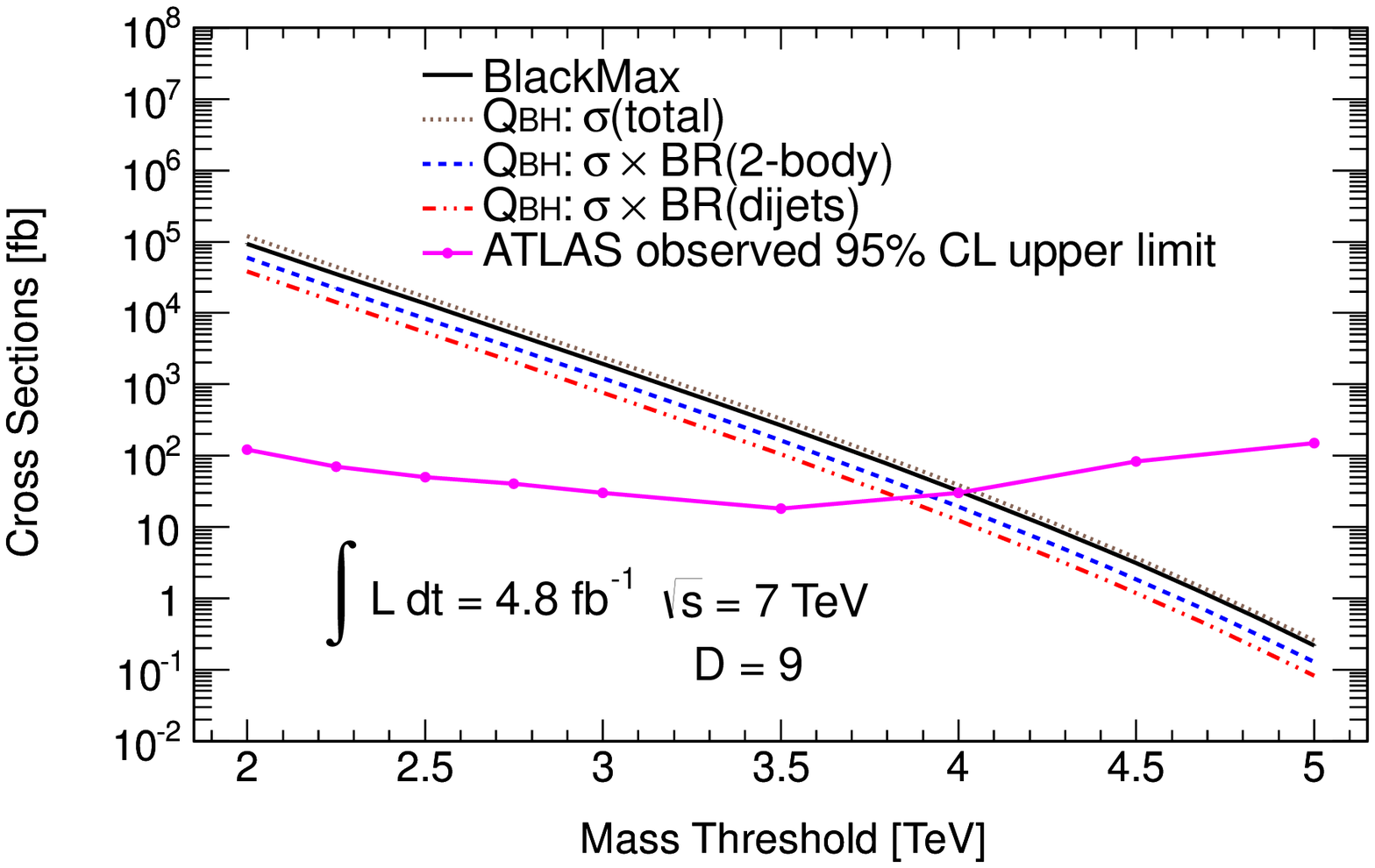}
\includegraphics[width=8.1cm]{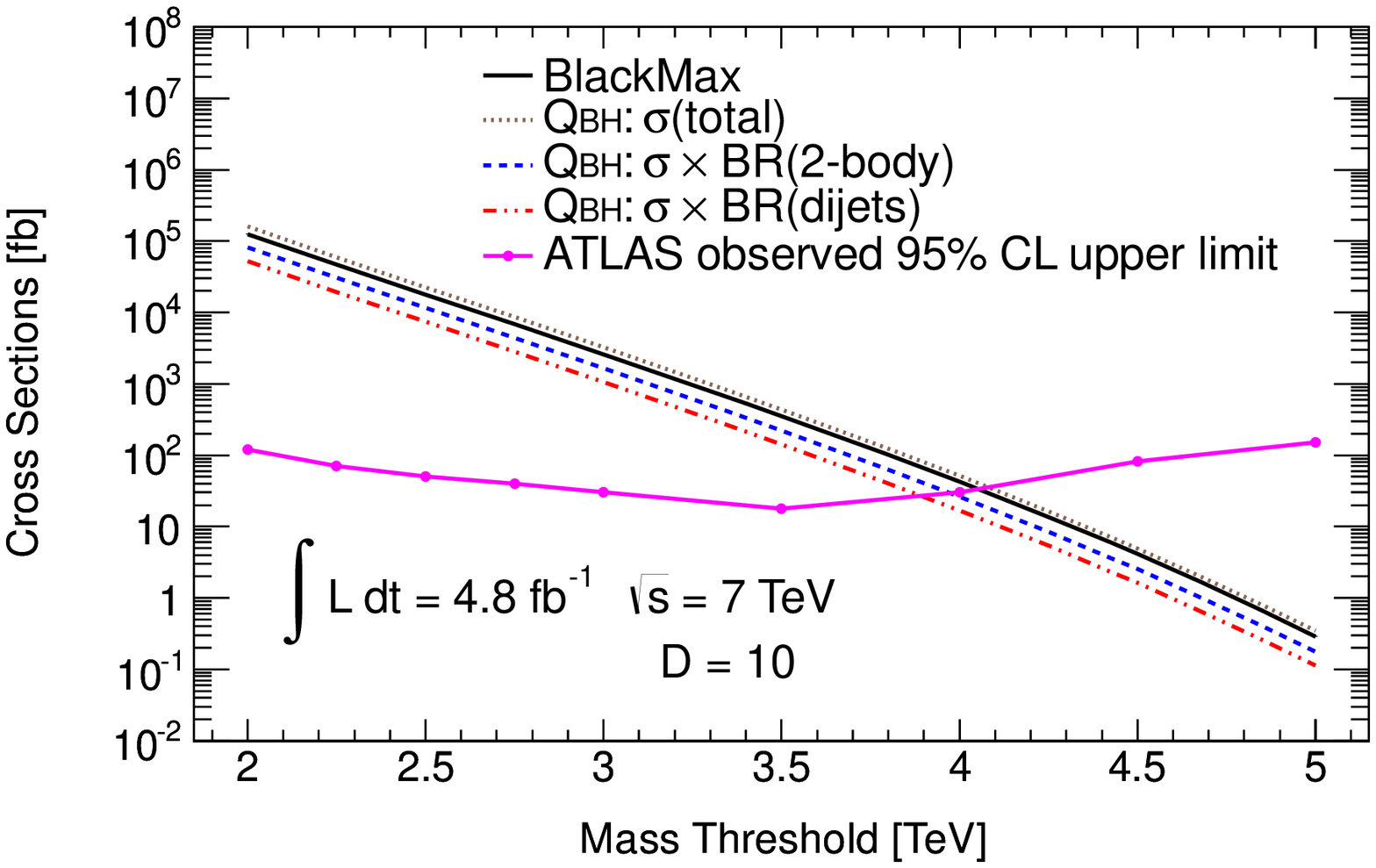}
\includegraphics[width=8.1cm]{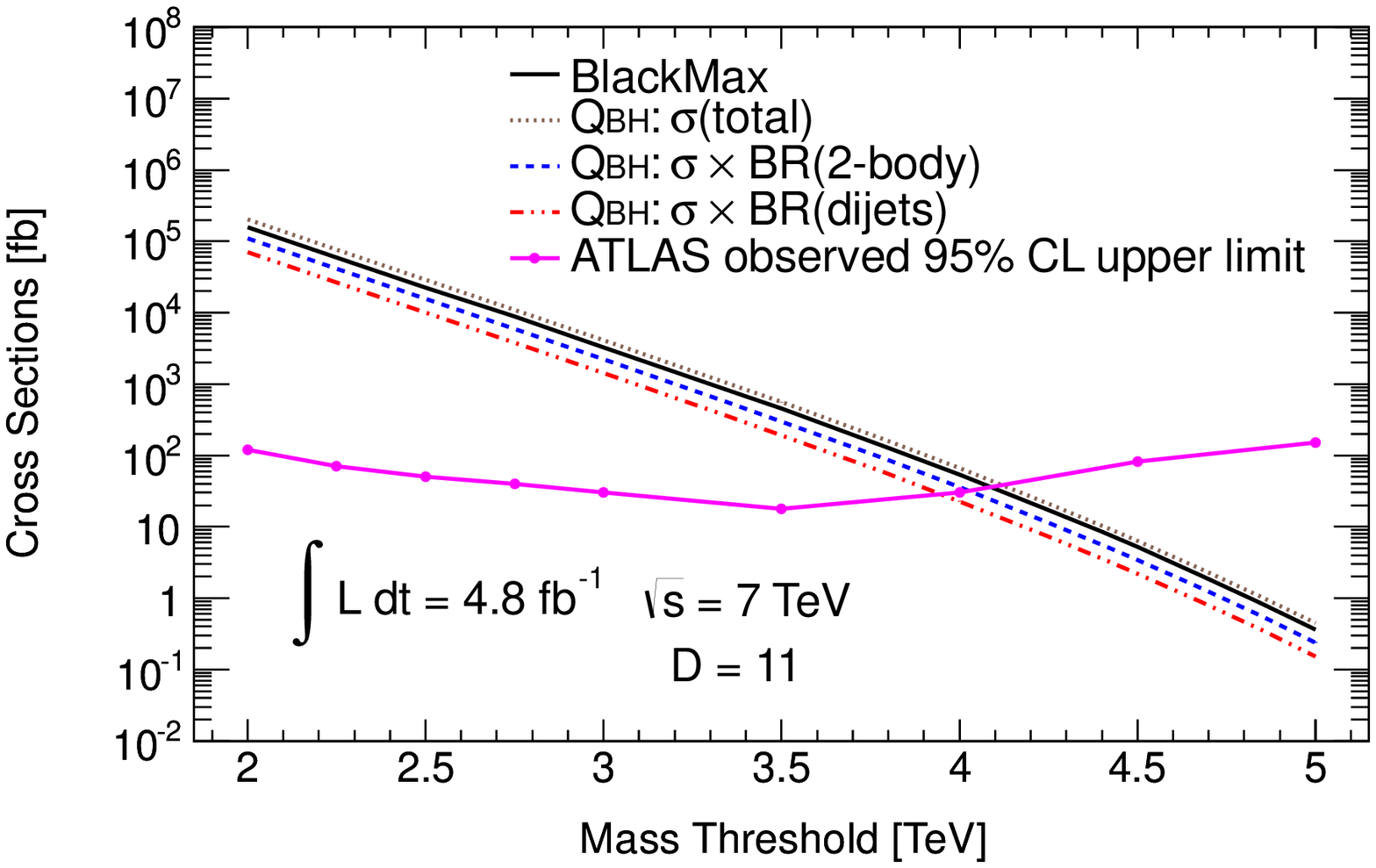}
\caption{Cross section versus mass threshold for different total number
of dimensions $D$ and branching ratio models.}
\label{fig2}
\end{figure}

The results of the different models for the branching ratio are shown in
Fig.~\ref{fig2}. 
The {\sc BlackMax} curve corresponds to the model used by the ATLAS
experiment.
The alternative interpretations are shown as the {\sc Qbh} curves.
The $\sigma$(total) curve presents the case of no two-body branching
ratio.
It assumes a black hole is always formed and always decays to two
partons which form two jets.
The two-body case implements the two-body branching ratio in
Eq.~(\ref{eq3}). 
Since not all decay particles from black holes are partons leading to
jets, the dijet curve shows the results when allowing for non-jet final
states. 
{\sc BlackMax} and the ATLAS analysis fail to take into consideration
the case where not all the final state particles appear as jets in the
detector. 

Table~\ref{tab1} shows the resulting lower mass limits for the different
model assumptions.

\begin{table}
\begin{center}
\begin{tabular}{ccccccc}\hline
$D$ & {\sc BlackMax} & \multicolumn{3}{c}{{\sc Qbh} [TeV]}\\\cline{3-5}  
& [TeV] & $\sigma$(total) & BR(2-body) & BR(dijets)\\\hline  
6  & 3.71 & 3.73 (+0.02) & 3.64 (-0.07) & 3.56 (-0.15)\\
7  & 3.84 & 3.88 (+0.04) & 3.76 (-0.08) & 3.67 (-0.17)\\
8  & 3.92 & 3.97 (+0.05) & 3.85 (-0.07) & 3.76 (-0.16)\\
9  & 3.99 & 4.03 (+0.04) & 3.91 (-0.08) & 3.83 (-0.16)\\
10 & 4.03 & 4.07 (+0.04) & 3.98 (-0.05) & 3.89 (-0.14)\\
11 & 4.07 & 4.11 (+0.04) & 4.02 (-0.05) & 3.95 (-0.12)\\
\hline
\end{tabular}
\caption{Lower limits at 95\% C.L. on the threshold mass versus total
number of dimensions $D$ for different quantum black hole branching
ratio models. 
The numbers in brackets are differences relative to {\sc BlackMax}.}
\label{tab1}
\end{center}
\end{table}

\section{Cross Section Uncertainties} \label{sec3}

Other different model assumptions, beside those of branching ratio, lead
to different cross sections and hence different limits on the threshold 
mass.
These are not restricted to models of quantum black holes but also apply
to the classical black hole models.
The scale used in the parton distibution functions can lead to cross
section differences.
{\sc BlackMax} uses the mass of the black hole, while {\sc Qbh} also allows the
inverse of the gravitational radius $1/r_g$ to be used.
Using the inverse gravitational radius can raise the limit on the
threshold mass by as much as 70~GeV.
Of course, different choices for the partons distribution functions can
give significant differences in cross section.
We do not consider these differences here but see
Ref.~\cite{Gingrich:2009hj} for some examples.

One can include a form factor based on the trapped surface calculation 
that is also used for classical black holes.
Including form factors raise the mass thresholds by as much as
200~GeV.

\begin{figure}
\centering
\includegraphics[width=8.1cm]{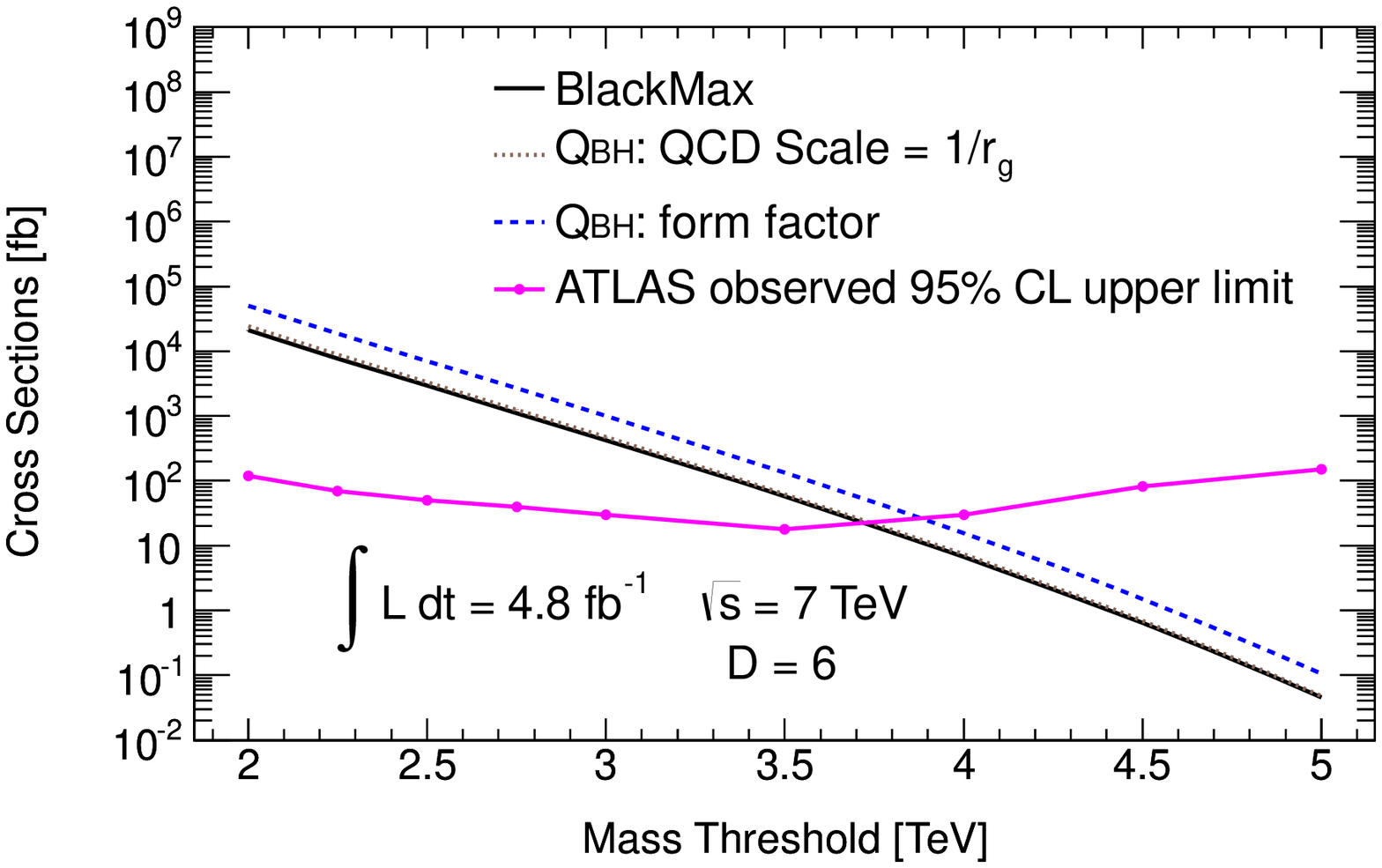}
\includegraphics[width=8.1cm]{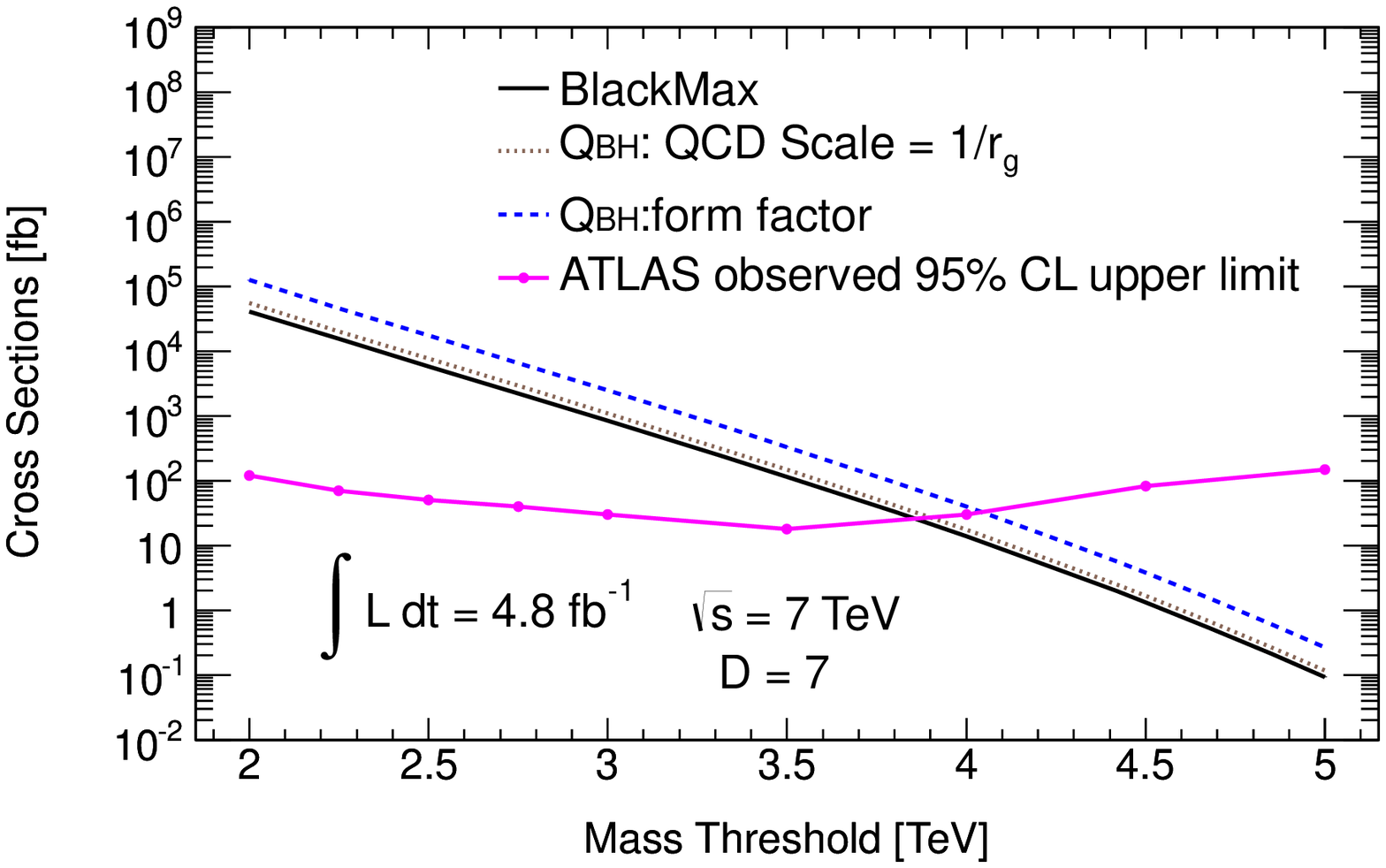}
\includegraphics[width=8.1cm]{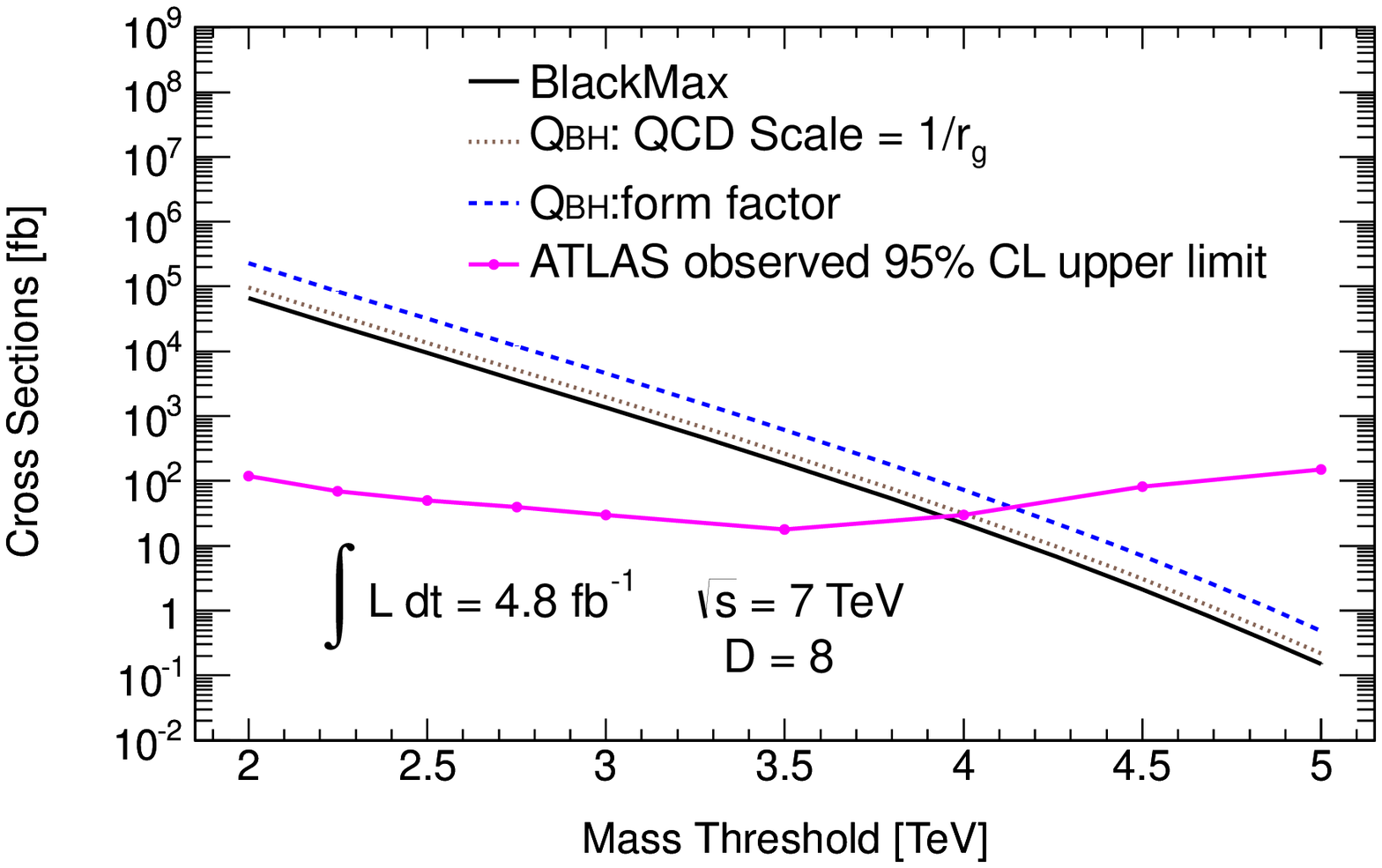}
\includegraphics[width=8.1cm]{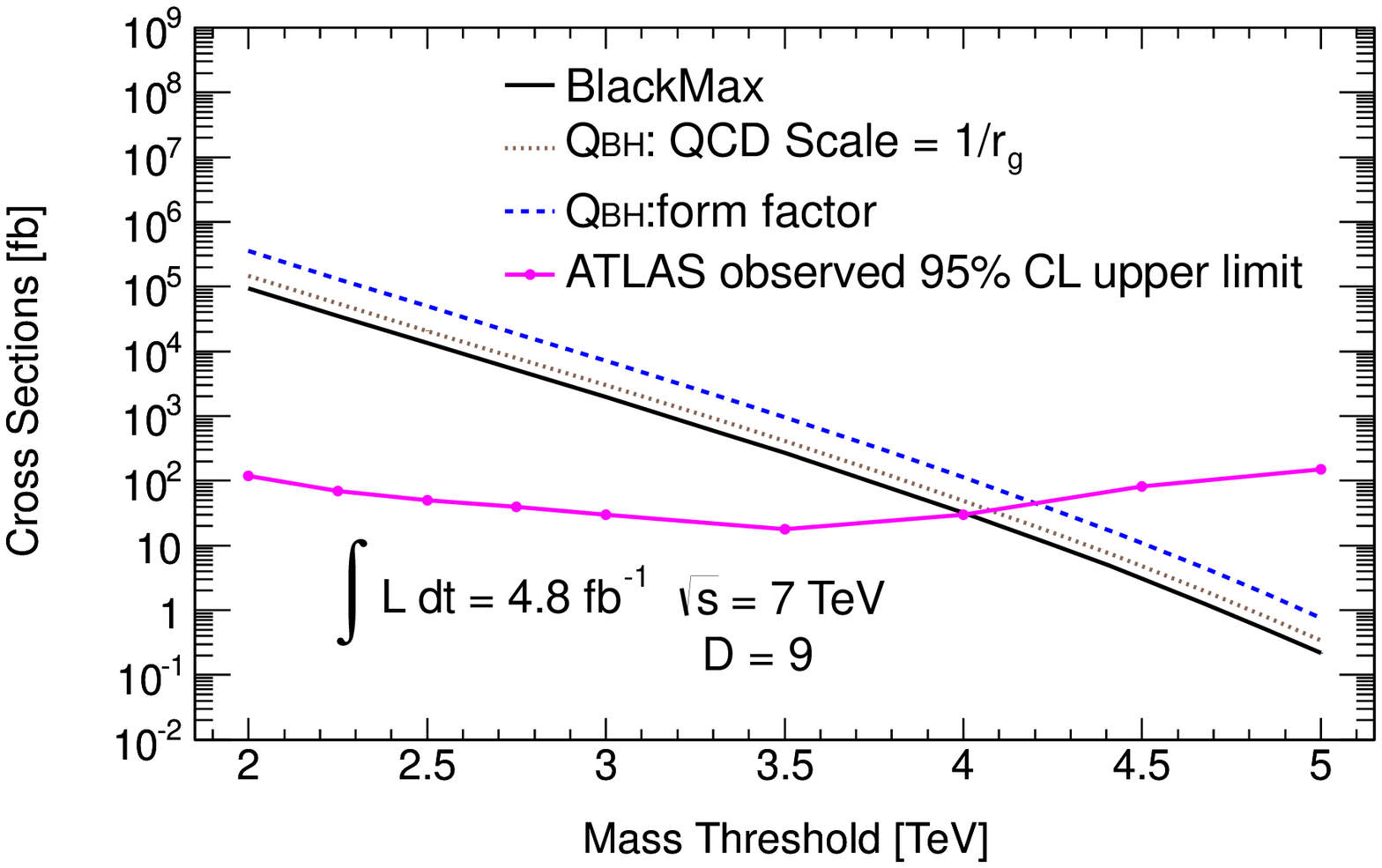}
\includegraphics[width=8.1cm]{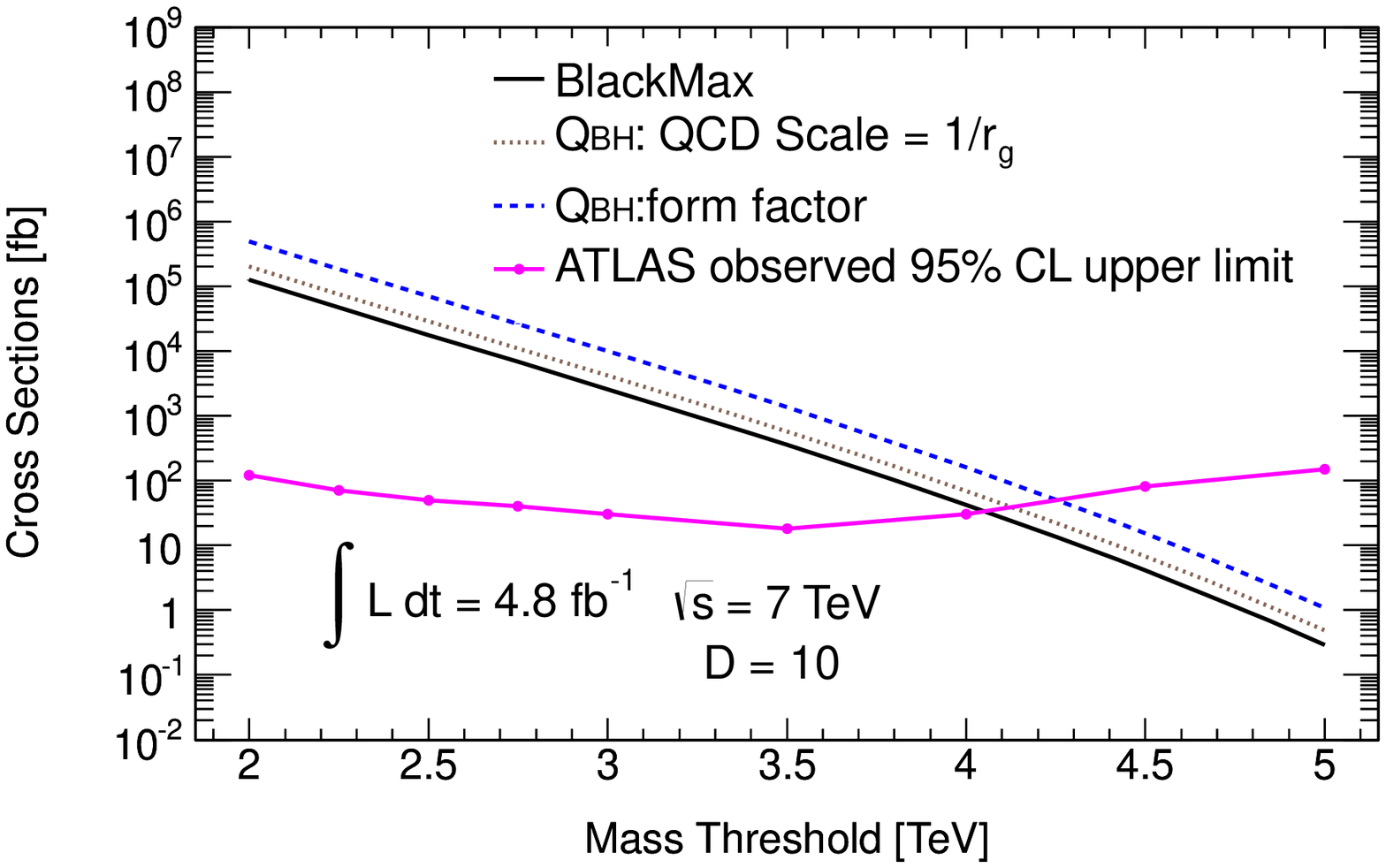}
\includegraphics[width=8.1cm]{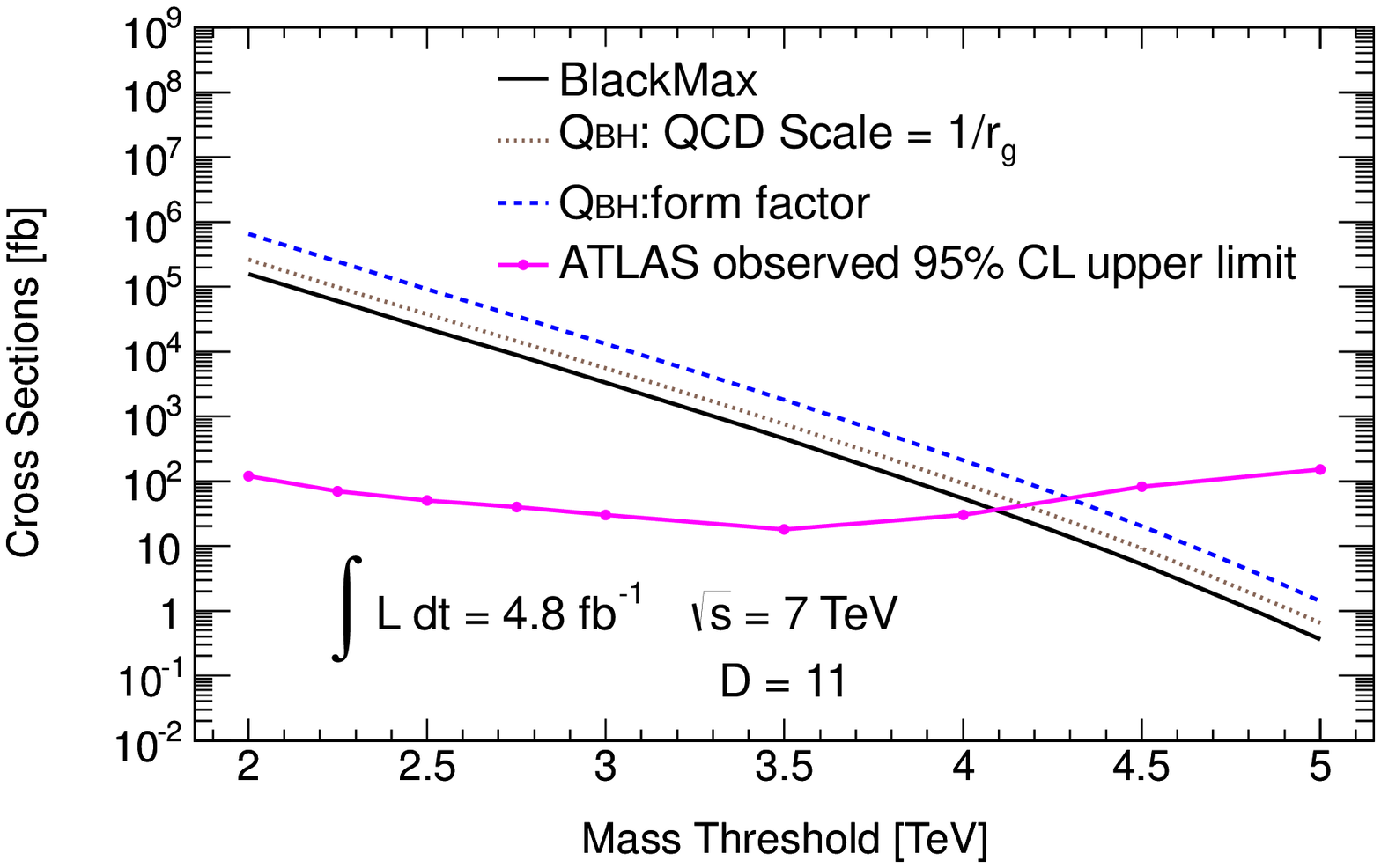}
\caption{Cross section versus mass threshold for different total number
of dimensions $D$ and for different cross section models.}
\label{fig3}
\end{figure}

The results of these two model assumptions are shown in Fig.~\ref{fig3},
and Table~\ref{tab2} shows the resulting lower mass limits for the
different model assumptions.

\begin{table}
\begin{center}
\begin{tabular}{ccccccc}\hline
$D$ & {\sc BlackMax} & \multicolumn{2}{c}{{\sc Qbh} [TeV]}\\\cline{3-4}  
& [TeV] & QCD scale $=1/r_g$ &
Form Factor\\\hline  
6  & 3.71 & 3.73 (+0.02) & 3.88 (+0.17)\\
7  & 3.84 & 3.90 (+0.06) & 4.04 (+0.20)\\
8  & 3.92 & 4.01 (+0.09) & 4.12 (+0.20)\\
9  & 3.99 & 4.07 (+0.08) & 4.18 (+0.19)\\
10 & 4.03 & 4.11 (+0.08) & 4.24 (+0.21)\\
11 & 4.07 & 4.15 (+0.08) & 4.28 (+0.19)\\
\hline
\end{tabular}
\caption{Lower limits at 95\% C.L. on the threshold mass versus total
number of dimensions $D$ for different quantum black hole cross section models.
The numbers in brackets are differences relative to {\sc BlackMax}.}
\label{tab2}
\end{center}
\end{table}

\subsection{Additional Limits}

\begin{figure}
\centering
\includegraphics[width=15cm]{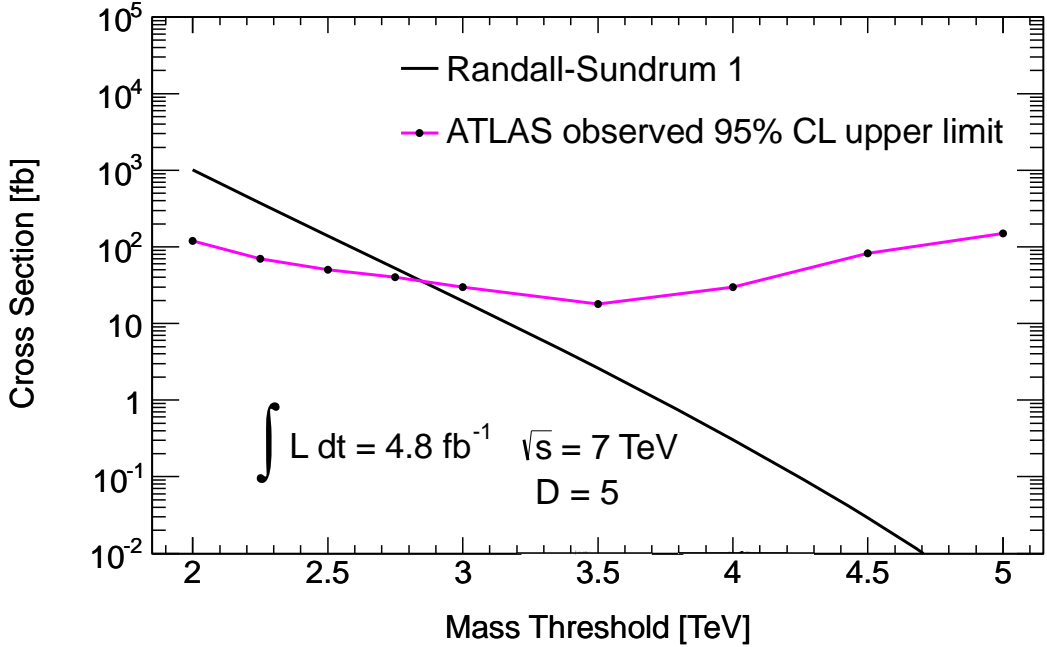}
\caption{Randall-Sundrum type-1 model cross section versus mass threshold.}
\label{fig4}
\end{figure}

It is possible to obtain additional information from the ATLAS data.
So far, ATLAS has considered only ADD-type models.
It is also possible to interpret the ATLAS data in terms of the
Randall-Sundrum type-1 model.
The result is shown in Fig.~\ref{fig4} and the mass threshold is
restricted to be above 2.84~TeV.

\begin{figure}[htb]
\centering
\includegraphics[width=15cm]{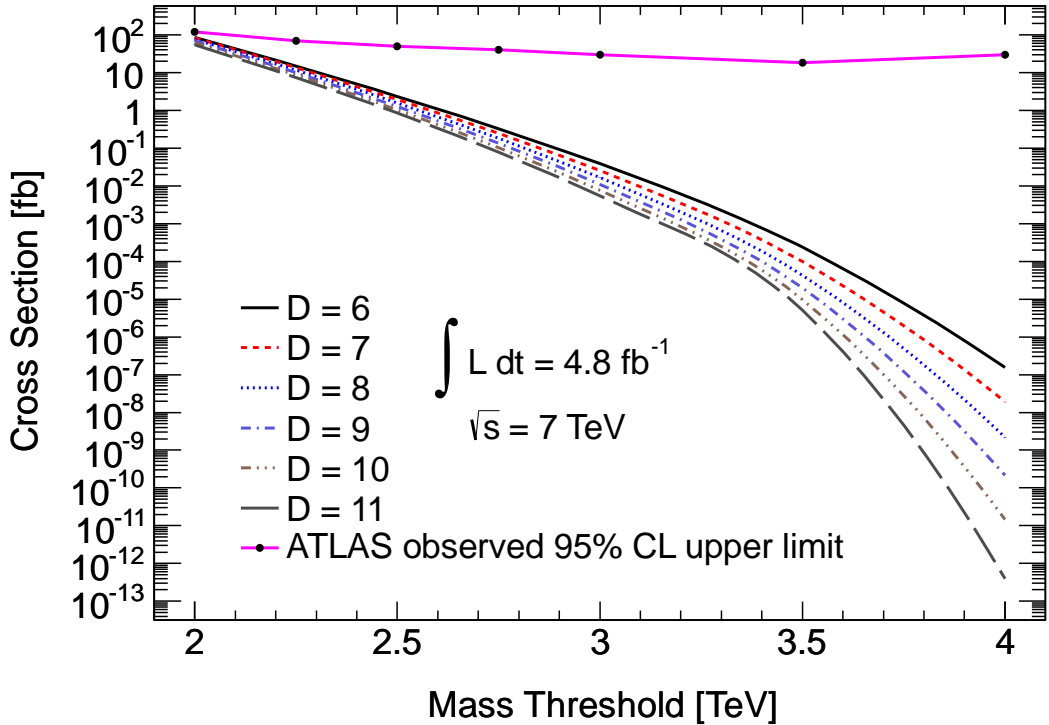}
\caption{Trapped surface cross section versus mass threshold.}
\label{fig5}
\end{figure}

The most significant effect on the cross section is the amount of energy
that goes into the formation of the black hole.
The classical cross section represents an upper bound, while the trapped
surface calculation gives a lower bound.
Figure~\ref{fig5} show the effect of using the trapped surface
calculation.
For this case, the ATLAS lower limits on the mass threshold are all
below 2~TeV. 

\section{Decay Uncertainties} \label{sec4}

Different models for the decay have no effect on the cross section,
other than the branching ratio, but may effect the experimental
acceptance.
Differences due to final state particle types in the branching ratio have
already been discussed.
ATLAS assumes an acceptance of 100\% when calculating the threshold mass
limits.
We believe the difference in acceptance due to different models for the
decay are negligible. 

It is thus unnecessary to preform a full detector simulation for all
the models and only the relative nomalization between the different
models is important. 
The $F_\chi(m_{jj})$ distriution observed by ATLAS in seeting the
limits would also be little changed from one model to another.

\section{Conclusions} \label{sec5}

Different model assumptions for the cross section raise the limits on
the mass threshold by at most 210~GeV.
Different treatments of the branching ratio reduce the limits on the
mass threshold by at most 170~GeV.
Thus, for models that assume a hard-disk cross section, we may assign an
approximate systematic error due to model dependence of about 5\% on the 
ATLAS results. 

\bigskip

This work was supported in part by the Natural Sciences and Engineering
Research Council of Canada.

\bibliographystyle{atlasnote}
\bibliography{dijets-qbh}
\end{document}